\documentclass[aps,prl,hyperref,showpacs,fleqn,amssymb,twocolumn]{revtex4}

\usepackage{graphicx}
\usepackage{amsmath}
\usepackage{bm}

\newcommand{\beq}{\begin{equation}}
\newcommand{\eeq}{\end{equation}}
\newcommand{\bD}{{\bf D}}
\newcommand{\bI}{{\bf I}}
\newcommand{\bQ}{{\bf Q}}

\newcommand{\bU}{{\bf U}}
\newcommand{\bsigma}{\bm{\sigma}}
\newcommand{\bSigma}{\bm{\Sigma}}
\newcommand{\bff}{{\bf f}}
\newcommand{\bk}{{\bf k}}
\newcommand{\br}{{\bf r}}
\newcommand{\bu}{{\bf u}}
\newcommand{\bx}{{\bf x}}
\newcommand{\bp}{{\bf p}}
\newcommand{\bn}{{\bf n}}
\newcommand{\bq}{{\bf q}}
\newcommand{\by}{{\bf y}}

\begin{document}
\title{Multiscale polar theory of microtubule and motor-protein
  assemblies}

\author{Tong Gao$^{1}$, Robert Blackwell$^{2}$, Matthew A. Glaser$^{2}$,
M. D. Betterton$^{2}$, Michael J. Shelley$^{1}$}
\affiliation{$^1$Courant Institute of Mathematical Sciences, New York
  University, New York, NY 10012 \\
  $^2$Department of Physics and
 Liquid Crystal Materials Research Center and
 Biofrontiers Institute
 University of Colorado, Boulder, CO 80309}

\begin{abstract}\noindent
  Microtubules and motor proteins are building blocks of
  self-organized subcellular biological structures such as the mitotic
  spindle and the centrosomal microtubule array. These same
  ingredients can form new ``bioactive'' liquid-crystalline fluids
  that are intrinsically out of equilibrium and which display complex
  flows and defect dynamics. It is not yet well understood how
  microscopic activity, which involves polarity-dependent interactions
  between motor proteins and microtubules, yields such larger scale
  dynamical structures. In our multiscale theory, Brownian dynamics
  simulations of polar microtubule ensembles driven by crosslinking
  motors allow us to study microscopic organization and stresses.
  Polarity sorting and crosslink relaxation emerge as two
  polar-specific sources of active destabilizing stress. On larger
  length scales, our continuum Doi-Onsager theory captures the
  hydrodynamic flows generated by polarity-dependent active
  stresses. The results connect local polar structure to flow
  structures and defect dynamics.
\end{abstract}
\maketitle

Nonequilibrium materials composed of self-driven constituents -- active
matter -- presents novel physics to understand and may
one day provide new technologies such as autonomously moving and
self-healing materials
\cite{Voituriez05,Wolgemuth08,ramaswamy10,Fielding11,Wensink12}. One
central example is mixtures of cytoskeletal filaments and molecular
motors, which are important for their ability to form self-assembled
cellular structures such as the mitotic spindle and cell
cortex. Reduced {\it in vitro} systems show that biofilament and motor-protein mixtures can form self-organized patterns, such as vortices
and asters, reminiscent of cellular structures
\cite{nedelec97,surrey01,schaller10}.  Recently, Sanchez {\it et al.}
\cite{sanchez12} synthesized mixtures of microtubules (MTs),
multimeric kinesin-1 motor complexes, ATP, and a depletant. In bulk,
extended MT bundles spontaneously form which continuously stretch,
bend and fracture, leading to large-scale flows. When condensed onto
an oil-water interface, the MTs form a nematically-ordered active
surface characterized by turbulent-like motions and motile
disclination defects.

Understanding reduced filament-motor systems is an important step
towards comprehending more complex active systems.  Therefore
theoretical studies have investigated aspects of MT and motor-protein
assemblies at different scales
\cite{nakazawa96,kruse00,Liverpool05,Woodhouse13,Head14}. Inspired by
the experiments of Sanchez {\it et al.} \cite{sanchez12}, Giomi {\it
  et al.}  \cite{giomi13,giomi14} and Thampi {\it et al.}
\cite{TGY2013,thampi14,thampi14b} have studied liquid crystal
hydrodynamic models driven by an apolar active stress \cite{simha02}.
While apolar models reproduce qualitative features of these
experiments, MTs have polarity and crosslinking motors move
directionally; hence, aligned MTs must have different interactions
than anti-aligned MTs, and activity-driven material stresses and
fluxes should reflect the polarity of these interactions. We
investigate this through multiscale modeling, first discovering two
separate microscopic sources of active and extensile stresses, one
induced by motor driven {\it polarity-sorting} of anti-aligned MTs,
and another from relaxation of crosslink tethers between polar-aligned
MTs. We formulate a Doi-Onsager model
\cite{doi88,saintillan08,KS2009,ESS2013} with fluxes and stresses reflecting
these effects, and use this to study the Sanchez {\it et al.}
interfacial experiments. Simulations show persistent folding flows and
defect birth and annihilation, arising from active stresses occupying
geometrically distinct regions. Having properly accounted for drag of
the bounding fluids, we find a well-defined characteristic
length-scale from linear theory which agrees well with feature sizes
in our simulations.

\begin{figure}
   \centering
      \includegraphics[scale=0.6]{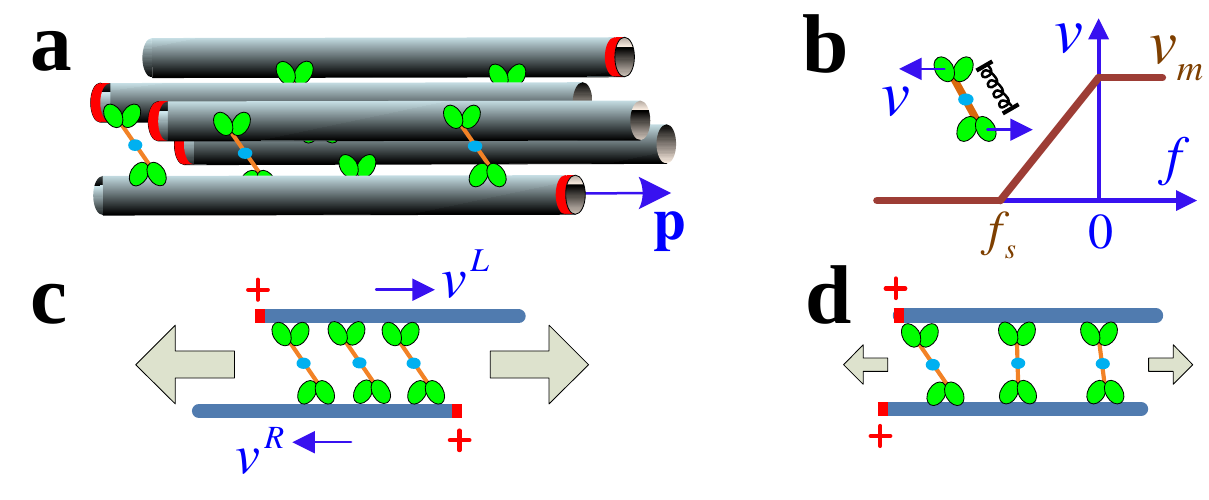}
      \caption{(a) Schematic of a cluster of polar-aligned and
        anti-aligned MTs, with their plus ends marked by red rings.
        Crosslinking motors walk on neighboring MTs at speed $v$, and
        (b) exert spring-like forces, with a linear force-velocity
        relation.  (c) An anti-aligned MT pair being polarity-sorted
        by active crosslinks. The left- (right-) pointing MT moves
        right (left) with velocity $v^L$ ($v^R$). (d) A polar-aligned MT
        pair upon which crosslink forces are relaxing due to the
        force-velocity relation. In both, the grey arrows characterize
        the magnitude of an induced extensile stress.}
   \label{schematic}
 \end{figure}

We outline the basic model in Fig. \ref{schematic}.  Every MT has a
 plus-end oriented director $\bp$, the same length $l$ and diameter
 $b$ (Fig.~\ref{schematic}a).  Nearby MTs are coupled by active
 plus-end directed crosslinks consisting of two motors connected by a
 spring-like tether. Motor velocities are controlled by a piecewise
 linear force-velocity relation (Fig.~\ref{schematic}b).
 For anti-aligned MTs (Fig.~\ref{schematic}c) the two motors move in
 opposite directions, stretching the tether to slide the MTs towards
 their minus-ends, which is termed {\it polarity sorting}
 \cite{nakazawa96}.  Conversely, for polar-aligned MTs the two motors
 move in the same direction, with little or no net sliding, and the
 retarding force on the leading motor causes stretched tethers to
 relax (Fig.~\ref{schematic}d).

\begin{figure}[!htbp]
    \centering
    \includegraphics[scale = 0.48]{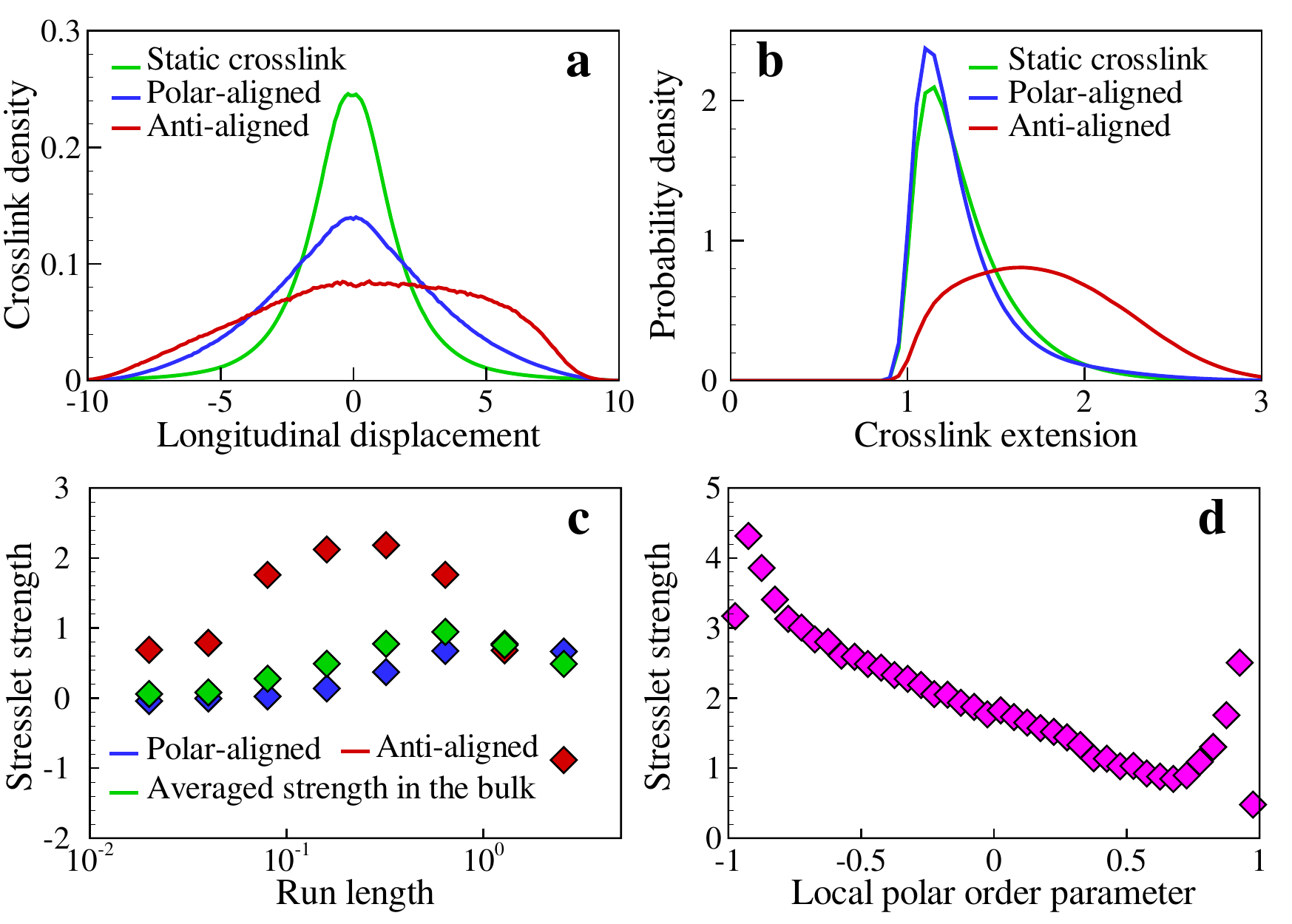}
    \caption{Results of the BD-MC particle simulations.  (a) Histogram
      of crosslink occupancy as a function of the particle pair
      longitudinal displacement $s_{ij}$. (b) Histogram of crosslink
      extension $r_c$. (c) Variation of extensile pair stresslet $S$
      (unit of force$\times$length) with motor run length $\ell$. (d)
      Typical variation of extensile pair stresslet with local
      polarity $m_i$. }
  \label{bdmc}
\end{figure}

{\it Microscopic model.}  We first perform 2D Brownian dynamics-Monte
Carlo (BD-MC) simulations of MTs driven by explicit motors with
binding/unbinding kinetics \cite{suppl}. The main purpose is to
quantify local MT pair interactions, with long-ranged hydrodynamics
neglected due to its high computational cost.  We represent MTs as
perfectly rigid rods, and assume a reservoir of ideal motors at fixed
chemical potential.  The motors bind to (unbind from) two filaments
simultaneously, and unbind immediately upon reaching the plus end of
either MT.  At equilibrium, the average number of motors crosslinking
MTs $i$ and $j$ is $\left\langle N_{ij} \right\rangle \sim \rho^2 \int
ds_{i} \int ds_{j} \exp{\left[-\frac{u_{c}(s_{i},s_{j})}{k_{B} T}\right]}$, where
$\rho$ is the linear binding-site density on a single MT, $u_c$ is the
quadratic potential for crosslinks extension, and $s_{i,j}$
parametrizes the MT arclength \cite{suppl}. The number of motors that
bind/unbind is sampled from a Poisson distribution with the correct
average number of events at each time interval so that the equilibrium
distribution is recovered for static crosslinks.  Bound motors are
inserted by first selecting pairs of MTs then sampling from the
appropriate bivariate normal distribution to choose motor endpoints.
The motor on each crosslink endpoint moves with a linear
force-velocity relation \cite{visscher99}:
$v=v_m\max(0,\min(1,1+f/f_s))$, where $f$ is the
magnitude of the crosslinking force, $v_m$ is the maximum
translocation velocity, and $f_s$ is the stall force. After the MC
cycle, we compute all the forces and torques from motors, short-range
repulsion, anisotropic local fluid drag by the solvent, and random
thermal forces, to evolve MT positions and orientations forward in
time \cite{Tao2005,suppl}.  The BD-MC simulations are
nondimensionalized using the length $b$, energy $k_B T$, and time
$\tau = D/b^2$, where $D$ is the diffusion coefficient of a sphere of
diameter $b$. Our model is similar to that of Head {\it et al.}
\cite{Head14}, but new in our work are algorithmic improvements for
handling crosslinks and neglect of filament elasticity that allow us
to simulate larger systems and measure the stress tensor.

{\it Extensile stress and its origins.}  Figure \ref{bdmc} illustrates
the long-time behavior of MTs in the BD-MC simulations (video S1).
For two MTs $i$ and $j$ with orientations $\bp_i$ and $\bp_j$ and
center-of-mass diplacement $\br_{ij}$, the longitudinal displacement
is $s_{ij}={1 \over
  2}\br_{ij}\cdot[\bp_i+\operatorname{sgn}(\bp_i\cdot\bp_j) \bp_j ]$.
For anti-aligned MT pairs ($\bp_i\cdot\bp_j<0$), $s_{ij}$ is negative
when the MT pair is contracting, and becomes positive when the MT pair
is extending (see Fig.~\ref{schematic}).  When crosslinks are static
or on polar-aligned MTs ($\bp_i\cdot\bp_j\geq 0$), the distribution of
crosslinks is symmetric about $s_{ij}=0$ (Fig.~\ref{bdmc}a).  However
for motors on anti-aligned MTs, the distribution skews toward positive
values of $s_{ij}$, yielding a bias in force generated by the motors
towards pair extension.

Motor motion alters the distribution of crosslink extension $r_c$
(Fig.~\ref{bdmc}b). The minimum $r_c \approx 1$ due to MT steric
interactions. For anti-aligned (polar-aligned) pairs, crosslink
relaxation shifts the distribution toward larger (smaller) extension.
The bulk material stress tensor is ${\bf \Sigma}_b = {{N k_{B} T} \over
  {V}} {\bf I} + {{1} \over {V}} \left\langle {\sum_i^N{\bf W}_i}
\right\rangle$ for $N$ interacting MTs in a volume $V$, with ${\bf
  W}_i = {1 \over 2} \sum_{j \neq i}^N {\bf r}_{ij} {\bf F}_{ij}$ the
single-MT virial tensor \cite{suppl,Allen87}.  Over a wide range of
motor parameters, the time-averaged bulk stress tensor $\bSigma_b$ is
anisotropic, with larger components in the average MT alignment
direction. Denoting the alignment direction by ${\hat\by}$, the stress
difference $\Sigma_b^{yy}-\Sigma_b^{xx}$ is positive, which
corresponds to an extensile stress. The stress difference can be
expressed as a sum of pair interactions, with each $ij$ pair
contributing a stresslet $S_{ij}$, prior to division by the bulk
volume. The average pair stresslet $S$ (green symbols in
Fig.~\ref{bdmc}c) increases with the motor run length $\ell$ up to a
maximum when the typical motor run length is the MT length. Here
$\ell= v_m/k_0 l$, the typical distance a motor travels during one
binding event, with $k_0$ a base binding rate of motors.  Increasing
$\ell$ further leads to decreasing $S$ because the motors rapidly move
to the ends of the MTs and unbind.

The extensile stress from anti-aligned pair interactions arises from
asymmetries during polarity sorting. If an MT pair begins sliding when
the two minus-ends touch and slide with a force proportional to pair
overlap until the two plus-ends meet, then the total extensile
stresslet would be zero.  Two effects break this symmetry. First, MTs
are unlikely to begin interacting exactly when their minus ends meet,
decreasing the range of negative $s_{ij}$ over which sliding occurs.
Second, more motors are bound on average during extension
(Fig.~\ref{bdmc}a). Fig.~\ref{bdmc}d shows a typical curve of $S$ as a
function of $m_i$, where the local polar orientational order parameter
$m_i$ varies between 1 (all neighboring MTs are polar-aligned) and
$-1$ (all neighbors are anti-aligned). A maximum for $m_i$ occurs near
$-1$ because polarity sorting is the dominant source of pair-wise
extensile stress.  As $m_i$ increases, $S$ drops with approximate
linearity, at least away from the two isolated peaks that close
examination shows are related to strong steric interactions of nearly
parallel MTs: nearly, but not exactly, parallel MTs experience
aligning torques due to crosslink-mediated attraction; the resulting
steric collisions tend to promote pair extension that increases the
extensile stress.

To understand the surprising and counterintuitive result that $S$
remains positive even for polar-aligned pairs, we consider crosslink
relaxation on perfectly parallel filaments.  When crosslinks are
active, the force of a longitudinally stretched crosslink opposes the
leading motor, slowing it, and pulls forward on the trailing motor.
This causes a slight but significant shift in the distribution of
crosslink extension toward smaller values relative to the
static-crosslink case (Fig.~\ref{bdmc}b). With crosslinking motors,
the crosslink-induced contractile stress along the MT alignment
direction is decreased, while there is no change in the transverse
stress induced by crosslinks. This leads to a net anisotropic
extensile stress in the alignment direction.  When varying system
parameters, we find that the extensile stresslet of polar-aligned MT
pairs is typically 2--5 times smaller than that of anti-aligned pairs.

{\it From microscopic to macroscopic models.}  To coarse-grain the
BD-MC simulation results, we introduce a
distribution function $\Psi(\bx,\bp,t)$ of MT center-of-mass positions
$\bx$ and polar orientation vectors $\bp$ ($|\bp|=1$), and describe
the particle dynamics in terms of the concentration $\Phi=\int_p\Psi$,
the polarity vector $\bq=\int_p\Psi\bp/\Phi$, the second-moment tensor
$\bD=\int_p\Psi\bp\bp$, and the tensor order parameter tensor
$\bQ=\bD/\Phi-\bI/d$, with $d=2$ or 3 the spatial dimension.

\begin{figure}[!htbp]
  \centering
   \includegraphics[scale = 0.46]{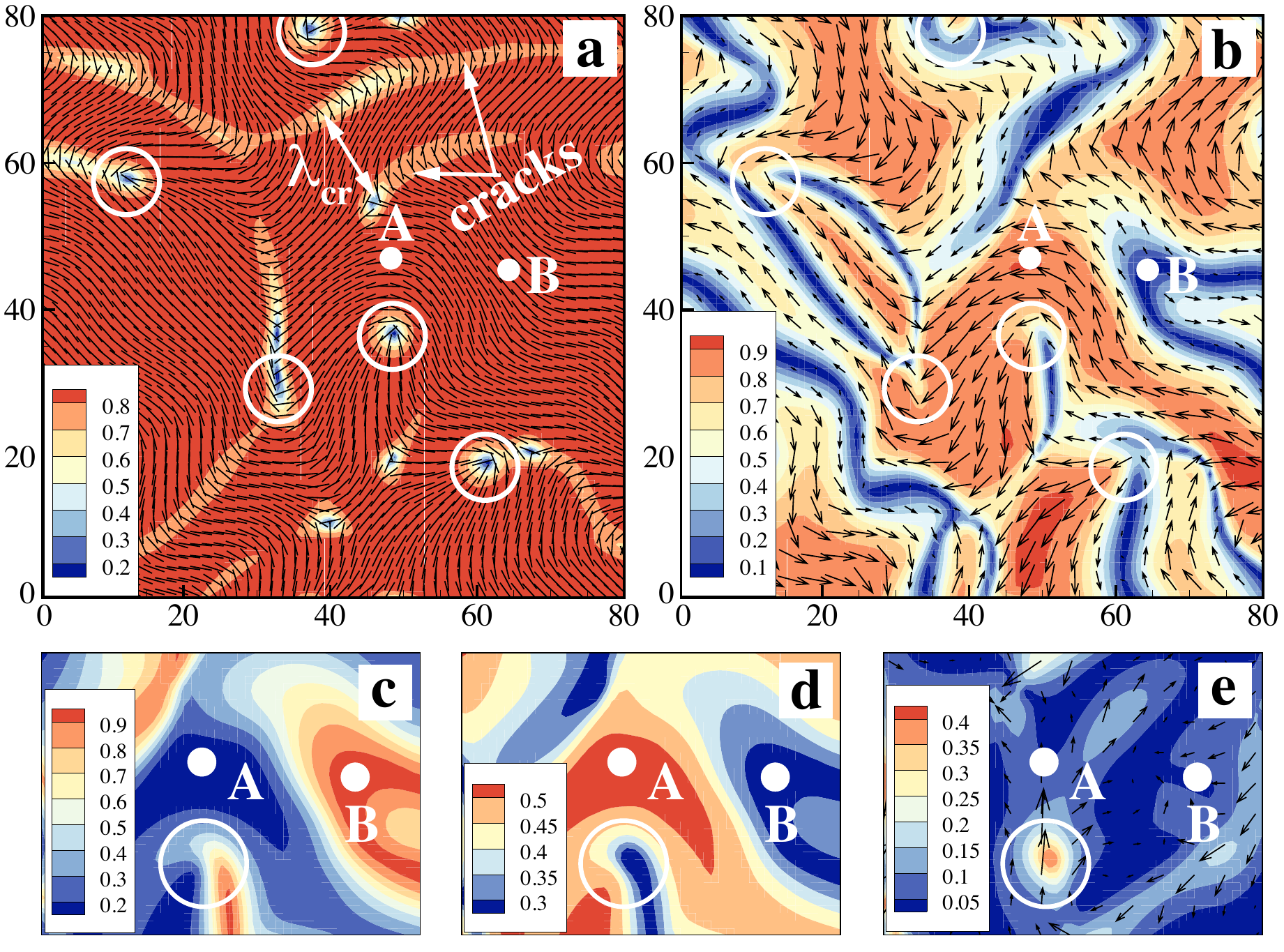}
   \includegraphics[scale = 0.47]{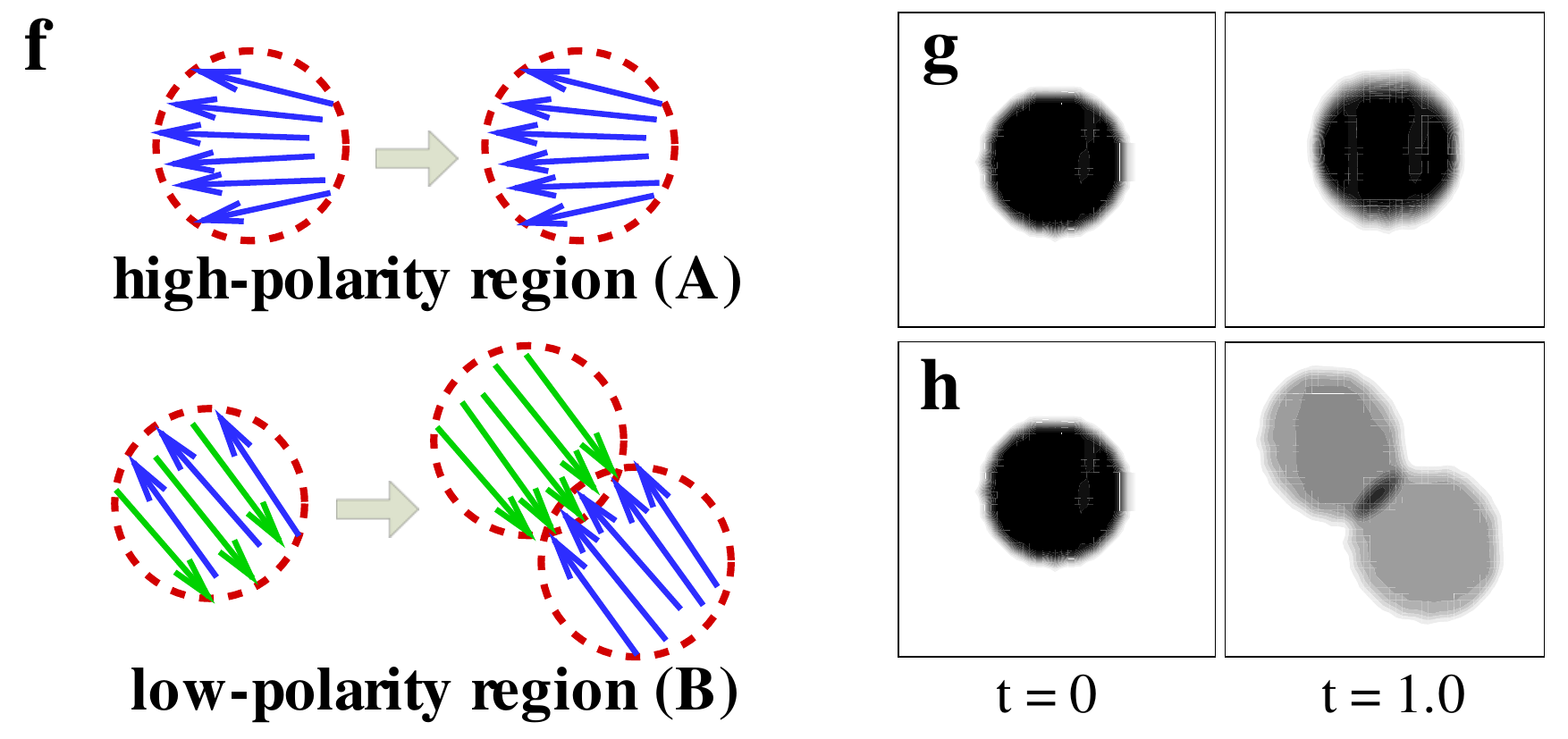}
   \caption{Snapshots of streaming MT nematics on a liquid-liquid
     interface. (a) The nematic director field $\bn$, superimposed on
     the colormap of the scalar order parameter (twice the positive
     eigenvalue of $\bQ$); $\lambda_{cr}$ is a calculated
     characteristic length between the cracks.  (b) The polarity
     vector field $\bq$ superimposed upon its magnitude.  (c,d)
     Polarity-dependent active stress magnitudes, showing principal
     eigenvalues of the active stresses (${\bf{\Sigma}}_{aa}$ in c,
     and ${\bf{\Sigma}}_{pa}$ in d).  (e) The vector field of the
     active force $\bff^a=\nabla \cdot \bSigma^a$ superimposed upon
     its magnitude. In (a-e), circular areas labeled A and B mark
     regions of high and low polarity, respectively.  Positions of
     $+1/2$-order defects are marked by circles.  (f-h) Predicted
     results of a photobleaching experiment of fluorescent MTs for a
     bleached spot in a region of high polar order (g, area A), and in
     a region of low polar order (h, area B).  Arrows represent MTs
     with arrowheads denoting plus ends. A dimensionless time is used
     with scale $b/\nu v_w$, where $\nu$ is the effective volume fraction.}
   \label{polarity}
\end{figure}

We first consider a nematically ordered local cluster of MTs
undergoing polarity sorting (Fig.~\ref{schematic}), with $n$ MTs
pointing rightwards (labelled $R$) and $m$ MTs pointing leftwards
(labelled $L$). When the motor ends move at a characteristic speed $v_w$, for an
anti-polar MT pair, a minus-end-directed sliding is induced.  Using
Stokesian slender body theory \cite{KR1976} we find the velocities of
the left- and rightward pointing MTs \cite{suppl}:
$v^L=\frac{2n}{{n+m}}v_{w},\quad {v^R} = - \frac{2m}{n+m}v_{w}$. This
expression shows that the speed of each population depends on how many
opposing MTs there are to pull against, with their drag as the anchor,
and their relative velocity fixed at $v^L-v^R=2v_w$. When considering a
more general orientation distribution, a similar calculation
\cite{suppl} yields ${\dot\bx} = \bq-\bp$ as the translational flux
for MTs.

Slender-body theory also yields the forces each rod exerts on the
fluid, and hence the induced ``extra stress'' tensor by polarity
sorting can be expressed in dimensional form as
$\bsigma_{aa}=\frac{{\eta} v_wl^2}{V_c}
\frac{\alpha_{aa}}{2}\frac{2mn}{m + n} \bp\bp$
\cite{suppl,batchelor70}. Here $V_c$ is the cluster volume, and
$\alpha_{aa}=s/l$ with $s$ the signed distance between the
centers-of-mass of the $\bp$ and $-\bp$ oriented subclusters. For the
extra stress due to crosslink relaxation, we lack a simple
first-principles model of polar-aligned MTs, though the number of
polar-pair interactions scales as $m^2+n^2$.  Given that the anti- and
polar-aligned stresses are of the same order (Fig.~\ref{bdmc}d) we
assume the form $\bsigma_{pa}=\frac{{\eta} v_wl^2}{V_c}
\frac{\alpha_{pa}}{2}\frac{m^2+n^2}{m+n}\bp\bp$. Thus we are able to
extract the (negative) values of $\alpha_{aa,pa}$ by comparing
the anti- and polar-aligned pair stresslet strengthes
($S_{aa,pa}=\frac{{\eta} v_w l^2 \alpha_{aa,pa}}{m+n}$, and $v_w$ is
taken as $v_m$) with the BD-MC simulations. Again, we construct the
dimensionless 3D extra stress from $\bD$ and $\Phi\bq\bq$ (i.e., the
simplest symmetric tensors quadratic in $\bp$) as $ \bSigma^a=
\bSigma_{aa}+\bSigma_{pa} =\frac{\alpha_{aa}}{2} ( \bD-\Phi\bq\bq )
+\frac{\alpha_{pa}}{2} ( \bD+\Phi\bq\bq )$. The first (second) term
captures active stress production via polarity sorting (crosslink
relaxation) and exactly reproduces the form of $\bsigma_{aa}$
($\bsigma_{pa}$).

We further account for particle rotation as well as steric
interactions, and couple MT motion with a background flow to study the
effect of long-range hydrodynamic interactions absent in the BD-MC
simulations through a continuum polar fluid model
\cite{doi88,saintillan08,Tjhung11,Forest13,ESS2013,suppl}. Since in
the Sanchez {\it et al.} experiments \cite{sanchez12} the active
material is confined to an interface between oil and water, we assume
a thin layer of suspension immersed in the bulk viscous liquid, and
close the system by solving the hydrodynamic coupling between the
surface flow $\bU$ and external fluid motions through a
velocity-stress relation in Fourier space: ${\hat \bU} = \frac{i}{2}(
\bI - {\hat\bk}{\hat\bk})({\hat\bSigma}^e{\hat\bk})$, where ${\bf{\hat
    k}} = {\bf{k}}/k$ is the normalized 2D wave-vector \cite{suppl}.

{\it Defects and polarity.} Assuming 2D periodic boundary conditions
and using a Fourier pseudo-spectral numerical method
\cite{saintillan08,suppl}, we simulated our model over long times.  In
regions of flow instability, we find persistently unsteady
turbulent-like flows that are correlated with continual genesis,
propagation, and annihilation of $\pm 1/2$ order defect pairs (see
videos S2--S5).  In Fig. \ref{polarity}a, the defects exist in regions
of small nematic order (dark blue), and are born as opposing pairs in
elongated ``incipient crack'' regions, qualitatively similar to the
structures found in both experiments and apolar models
\cite{giomi13,giomi14,TGY2013,thampi14,thampi14b}. Moreover, as shown
in panel (b), the polarity field develops considerable spatial
variation with regions of high and low polar order $|\bq|$ (video S5).
The two active stresses vary in strength depending on the local
polarity --- the polar-aligned (anti-aligned) stress is large in
regions of high (low) polar order (panels (c) \& (d)) --- and hence
are largest respectively in their complementary regions. The circles
in Fig.~\ref{polarity}b encircle $+1/2$-order defects, showing the
sharp variation of the polarity field around them, with the gradients
of active stresses there yielding large active force shown in panel
(e).

We further simulated the results of a photobleaching
experiment in which a circular region is exposed to high-intensity
laser light to inactivate the fluorescent molecules on the
corresponding MTs \cite{AKSEW1976} (Figs.~\ref{polarity}f-h). In a
small high-polarity region (marked A in Fig.~\ref{polarity}), little
or no polarity sorting occurs, and the photobleached spot
remains approximately circular (Fig.~\ref{polarity}f top, g)
over longer times. In a low-polarity
region of high nematic order (marked B in Fig.~\ref{polarity}), strong
polarity sorting of anti-aligned MTs causes a photobleached spot to
separate into two lobes, showing decreased
bleaching. This type of experiment probes the local polarity field,
and hence the origins of active stress.

\begin{figure}
  \centering
   \includegraphics[scale = 0.45]{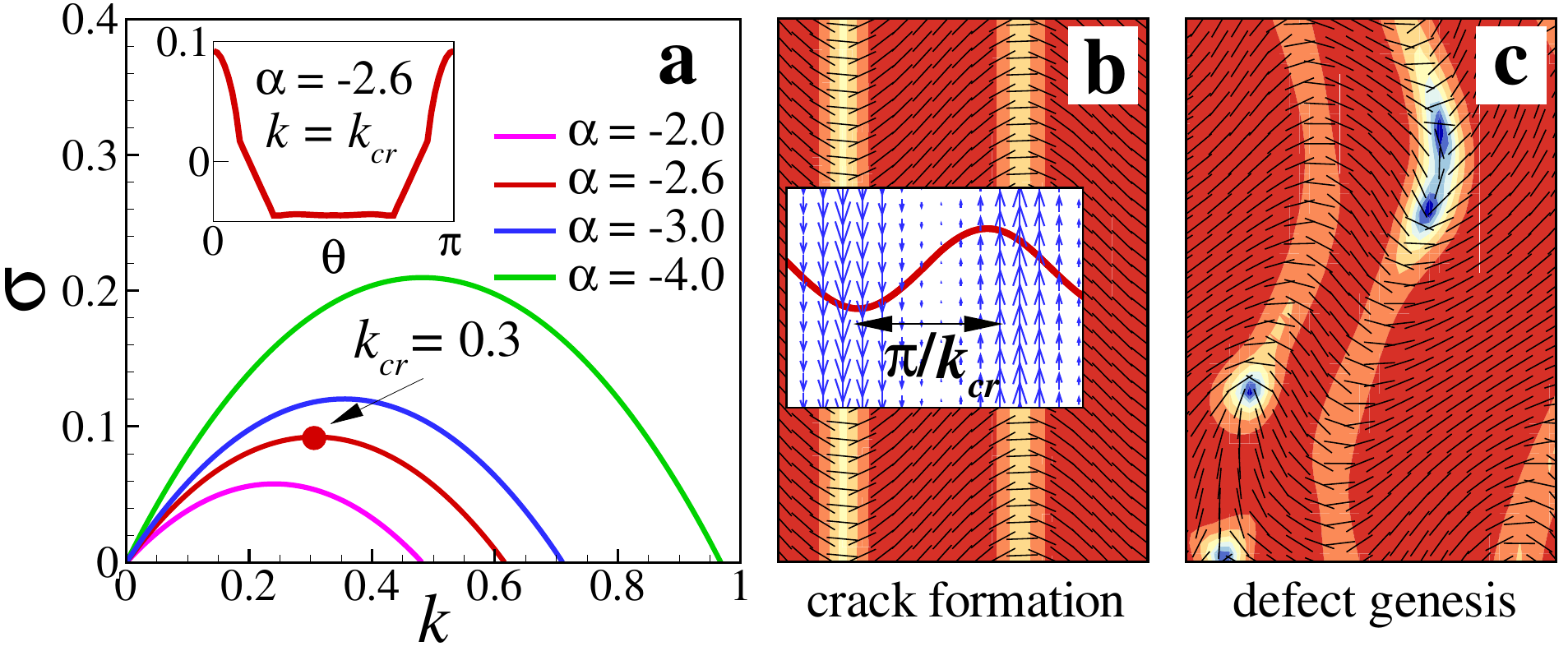}
   \caption{ Linear stability analysis (a) and nonlinear simulation
     (b,c) for strongly anti-aligned MTs. (a) The real part of the
     growth rate as a function of wave-number $k$ for several values of
     $\alpha$ ($\alpha = \alpha_{aa} + \alpha_{pa}$). Here $k_{cr}$
     corresponds to a maximum growth rate. Inset: real part of the
     growth rate as a function of wave-angle $\theta$ when fixing $k =
     k_{cr}$. (b) Crack formation. Inset: the fluid velocity vector
     field (blue) and the eigenmode (red line) associated with $k_{cr}$.
     (c) Genesis of defects at late times.}
 \label{instability}
\end{figure}

{\it Hydrodynamic instabilities and characteristic length.}  A key
observation in our simulations as well as other active fluid systems
\cite{thampi14,Giomi11,Giomi12a,zhou14} is that defect pairs are
generated along elongated cracks that themselves develop from regions
of high polar order. Here we performed a linear stability analysis for
a nematically ordered homogeneous base-state \cite{suppl}.  As shown
in Fig.~\ref{instability}a, the analysis reveals a wave-number of
maximal growth, $k_{cr}$, along this direction, with $k_{cr}$ growing
with approximate linearity in $\alpha=\alpha_{aa}+\alpha_{pa}$. Also
the plane-wave vector of maximal growth is aligned with the nematic
director ($\theta=0$ in Fig.~\ref{instability}a inset).  By contrast,
when solving a Stokes equation forced by a bulk stress, the
velocity-stress relation in Fourier space becomes
${\hat\bu}=\frac{i}{k}(\bI-{\hat\bk}{\hat\bk})({\hat\bSigma}^e{\hat\bk})$.
Comparing to our surface model, the factor of {\bf $k$} in the
denominator profoundly changes the nature of system stability, giving
that maximal growth occurs at $k=0$ for the bulk model, and so not
producing a characteristic length-scale
\cite{saintillan08,private}. (This was also noted by \cite{Leoni10} in
their study of swimmers confined to immersed thin films, while
\cite{thampi14c} show that adding substrate friction changes
length-scale selection in 2D active nematic models.)

Figure~\ref{instability}b shows the nonlinear results of this linear
instability. A series of cracks form along $\hat{\bf y}$, associated
with moving fluid jets and bending of nematic field lines. In panel
(b) inset, the spatial variations of the velocity field are in
excellent agreement with the velocity eigenmode associated with
$k_{cr}$ for the linearized system. The distance between these cracks
matches the half-wavelength, i.e., $\lambda_{cr} = \pi \, /k_{cr}$,
which is in fact representative of the characteristic of the full
dynamics of motile defects (Fig. \ref{polarity}a). At late times,
panel (c) shows that these cracks lose stability, and eventually
``break'' to form defect pairs.

{\it Discussion.} We have explored other aspects of our model system.
For example, when turning off hydrodynamics in our kinetic model, we
find polar lanes emerging as in our BD-MC model. This arises from a
slow instability (consistent with the BD-MC model) when compared with
hydrodynamic instabilities. We find that either active stress
({\it aa} or {\it pa}) taken individually will produce qualitatively
similar flows and defect dynamics. Hence, the qualitative nature of
the large-scale dynamics does not by itself isolate the precise
origins of a destabilizing stress. An interesting aspect of our BD-MC
study is that active stresses are extensile, which is very different
from the contractility observed in actin-myosin gels
\cite{bendix08}. This is likely related to the rigid MTs being in a
nematically ordered state. While we applied our multiscale polar model
to study experiments of synthesized active fluids, similar but more
elaborated models might serve as a principled basis from which to
study biological systems such as the eukaryotic mitotic spindle.

{\it Acknowledgements.} We thank D. Chen and D. Needleman
for useful discussions. This work was funded by NSF grants
DMR-0820341 (NYU MRSEC: TG, MS), DMS-0920930 (MS), EF-ATB-1137822
(MB), DMR-0847685 (MB), and DMR-0820579 (CU MRSEC: MG); DOE grant
DE-FG02-88ER25053 (TG, MS); NIH grant R01 GM104976-03 (MB, MS); and
the use of the Janus supercomputer supported by NSF grant CNS-0821794.

\end{document}